\documentstyle[aps,pre,psfig,floats]{revtex}
\begin{document}                             
\newcommand{\C}{{\cal C}}
\newcommand{\G}{{\cal G}}
\newcommand{\EA}{{\rm EA}}
\psfigurepath{/users3/guyh/tmp/eps:/users3/guyh/spin_glasses/gs3d/figs}

\title{Correct extrapolation of overlap distribution in spin glasses}
\author{Guy Hed$^1$, Alexander K. Hartmann$^2$ and Eytan Domany$^1$\\[2mm]}
\address{$^{1}$ Department of Physics of Complex Systems,
    Weizmann Institute of Science, Rehovot 76100, Israel\\
    $^{2}$ Institut f\"ur Theoretische Physik, Universit\"at G\"ottingen,
    Bunsenstr. 9, 37073 G\"ottingen, Germany }
\date{\today}
\maketitle

\begin{abstract}
We study in $d=3$ dimensions the short range Ising spin glass
with $J_{ij}=\pm1$  couplings at $T=0$.
We show that the overlap distribution is non-trivial in the limit
of large system size.
\end{abstract}

\vspace{20pt}
\section{Introduction}
There has been an upsurge of interest in the Edwards-Anderson model \cite{EA}
of the short-range Ising spin glass with binary couplings,
\begin{equation}
{\cal H} = \sum_{i<j} J_{ij} S_i S_j \qquad \qquad J_{ij}=\pm1
\label{eq:model}
\end{equation}
in $d=3$ dimensions. 
Several papers addressed the following well-posed question: 
\begin{quote}
{\it is the overlap
distribution $P(q)$ of this model, measured  at $T=0$, trivial or non-trivial
in the thermodynamic limit?}
\end{quote}
A trivial $P(q)$ consists of a single delta
function, $P(q) = \delta(q-q_{\EA})$; a non-trivial $P(q)$ has non-vanishing
support also for $0< q <q_{\EA}$.
Claims were made to the effect that the nature of $P(q)$ 
bears on the validity of the droplet picture (DP) \cite{droplet}
versus the scenario \cite{RSB} based on mean-field theory (MF) \cite{SK}. 
Our understanding is that 
a non-trivial global\footnote{
We assume that $P(q)$ is measured for an entire system,
whose size is then extrapolated to $L \rightarrow \infty$.}
$P(q)$ is consistent with both \cite{Huse87}. 
In this communication we do not take sides in the DP vs MF controversy;
rather, we address the well-defined technical question posed above. 
\begin{quote}
{\it Our conclusion is that $P(q)$ is non-trivial at $T=0$.}
\end{quote}
Berg et al \cite{Berg94} addressed the issue directly by generating 
ground states ${\mathbf{S}}^\mu=(S^\mu_1,S^\mu_2,...S^\mu_N)$,
for 512 realizations $\{ J \}$
of systems with sizes $L=4,6,8$ (and for 7 realizations of
$L=12$). For each $\{ J \}$ they computed the overlap distribution 
function $P_J (q)$, where the overlap 
$q^{\mu\nu}=(1/N)\mathbf{S}^\mu \cdot \mathbf{S}^\nu$ is calculated 
between all pairs of ground states $\mu,\nu$. They 
studied the function
obtained by averaging over all realizations, $P(q) = [P_J(q)]_J$. 
In particular, they evaluated $P(0)$; the second moment of the distribution
$\sigma^2(q)$, and the quantity $x_{1/2}$, where $x_a$ is defined by
\begin{equation}
x_{a}=2 \int_0^{a} P(q) d q
\label{eq:xa}
\end{equation} 
If $P(q) \rightarrow \delta(|q|-q_{\EA})/2$ for large $L$, all these quantities
should extrapolate to zero 
(provided one uses $a<q_{\EA}$). 
Berg et al found that
all three quantities decrease as $L$ grows; they could, however, fit the 
data to 
$ L^{-y}$, with $y=0.72\pm .12$, as well as to $A+BL^{-3}$, indicating 
consistency with 
extrapolation to both vanishing and non-vanishing limiting values.

Hartmann \cite{Hartmann3d} also studied the size-dependence of $x_{1/2}$
and found that it behaves as $L^{-y}$ with $y=1.25 \pm .05$, indicating
a trivial $P(q)$; the same conclusion was
reached by Hatano and Gubernatis \cite{Hatano00}
who studied $P(0)$ at finite temperatures.
Krzakala and Martin \cite{KMbinary}
presented  arguments that also support a trivial $P(q)$.
Finally, very recently Palassini and Young \cite{PYbinary}
evaluated $P(q)$ for a sequence of
temperatures and sizes $L=4,6,8,10$. They evaluated $x_{1/2}$ as function
of $L$ and $T$ and demonstrated that the data are consistent with a scaling
form. According to their scaling, for fixed $T>0$ and sizes $L \gg L_c(T)$, 
$x_{1/2}$ goes to a constant, $x^\infty(T) \propto T$; 
hence they find that  $P(q)$ is non-trivial at $T>0$ 
and trivial at $T=0$. 

\section{Outline of Strategy}

We will argue now that all the studies mentioned  measured a compound 
quantity, $x_{1/2}$, which is the sum of two parts;
one which is relevant to the question asked, and another
which is irrelevant. Furthermore, for some of  the sizes studied,
the irrelevant
part is as large as the relevant one. We will show how can one isolate the
relevant part, and present the results obtained when this is done. 
These results indicate that $P(q)$ is non-trivial at $T=0$.

\begin{figure}[t]
\centerline{\psfig{figure=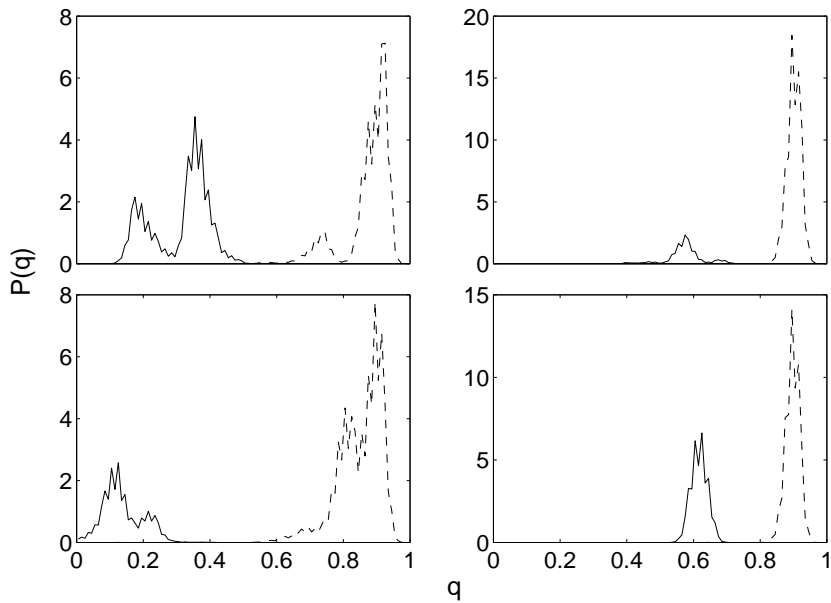,width=11cm}}\vspace{2mm}
\caption{The state overlap distribution $P(q)$ for four different
realizations $\{J\}$ for system size $L=8$. Each distribution is divided to
its components. 
The partial distribution $\tilde P_J^o(q)$ (see text), is represented
by a solid line. The rest of the distribution, $P_J(q)-\tilde P_J^o(q)$,
which includes $P_J^i(q)$, is represented by a dashed line.}
\label{fig:pqj}
\end{figure}

For a particular realization  $P_J(q)$ has, at $T=0$, 
the typical form presented in Fig \ref{fig:pqj}. It has a large peak
centered at some $q_0 \approx q_{\rm EA}$, and one or more smaller peaks.
The largest peak is due to the overlap of pairs of states  that belong to
the {\it same} pure state. Denote the overlap distribution of such
pairs by $P_J^i(q)$.  The other peaks, at lower $q$, are due to the overlap
between states that belong to two {\it different} pure states.
The corresponding overlap distribution is $P_J^o(q)$ and we have
\begin{equation}
P_J(q)=P_J^i(q)+P_J^o(q) \qquad {\rm and} \qquad  P(q)=P^i(q)+P^o(q)
\end{equation}
where the second equation is the average of the first over all realizations.
Hence we can write 
\begin{equation}
x_a=x_a^i + x_a^o= 2\int_0^a P^i(q) + 2\int_0^a P^o(q)
\end{equation}
Irrespectively of whether $P(q)$ is trivial or non-trivial,
one expects that the width
of $P^i(q)$ decreases with increasing size, since
$P^i(q) \rightarrow\delta(|q|-q_{\EA})/2$ as $L \rightarrow \infty$. 
Therefore as $L$ increases, 
the integral $x_a^i$ decreases towards 0;  
\begin{equation}
x_a^i \approx L^{-y_i} 
\label{eq:decay}
\end{equation}
On the other hand, the behavior of $P^o(q)$ (and $x_a^o$) {\it does} 
distinguish a trivial $P(q)$  from a non-trivial one;  in the first
case $P^o(q) \rightarrow 0$, while in the 
non-trivial case $P^o(q)$ and $x_a^o$ do not vanish 
as $L \rightarrow \infty$. 
We believe that  previous analysis was hindered by the lack of ability to
decompose $P(q)$ and $x_{1/2}$ into its two constituent parts; a method that 
we developed recently  enables us to perform this task.
We describe below how we can use a very recently developed method~\cite{PRL} 
to identify unabiguously, for a large majority of the realizations,
a partial distribution ${\tilde P}_J^o(q)$, which is a lower bound to 
$P_J^o(q)$. For realizations in which this identification is ambiguous we set 
${\tilde P}_J^o(q)$
to zero;  averageing  yields $\tilde P^o(q)=[{\tilde P}_J^o(q)]_J$, which is a
lower bound
on $P^o(q)$.
The corresponding lower bound on $x^o_a$ is given by
\begin{equation}
\tilde x^o_a=2\int_0^a \tilde P^o(q) dq \;,
\label{eq:xtilde}
\end{equation} 
%

We found that the rate of convergence of 
${\tilde P}^o(q)$ to its limiting
large-$L$ form is non-uniform; for the sizes studied, convergence 
(with increasing $L$) is much slower,
and statistical errors are much larger in the interval $0 \leq q \leq 0.5$ than
in $0.4 \leq q \leq 0.7$. Hence we base our analysis on the latter 
interval, calculate
\begin{equation}
x^*=2\int_{0.4}^{0.7} P(q) dq
\label{eq:x*}
\end{equation} 
and show that it
approaches a 
non-vanishing limit as  $L \rightarrow \infty$.

\section{Decomposing $P_J(\lowercase{q})$}

Our method has been presented in~\cite{PRL}, together with results
obtained for the model (\ref{eq:model}). Full details of the method 
are given in~\cite{Gauss}; here we give a brief summary of the main 
ingredients.

We have shown that an unbiased sample~\cite{PRL} of $M$ ground states
breaks naturally into two large groups, $\C$ and $\bar \C$. The states
of the two sets are related by spin reversal. For a large majority of
realizations the set $\C$ also breaks into two natural subsets,
$\C_1$ and $\C_2$. By natural we mean that  the overlap between 
two states that belong to the same group (say $\C_1$) is significantly 
larger than between two that belong to two different groups. 
This suggest that the states in these clusters belong to different pure
states, separated by free energy barriers. These barriers consist of
correlated spin domains $\G_1$ and $\G_2$, which flip collectively
when we move from a state in one cluster to a state in another cluster.
The spins that belong to $\G_1$ are reversed in at least $95 \%$ of the pairs
of states $\mu \in \C$ and $\nu \in {\bar \C}$. Similarly, the second largest
domain $\G_2$ contains those spins that flip in $95 \%$ (or more) of the
times we pass between pairs  states, with one member in $\C_1$ and the other
in $\C_2$. 

The domains $\G_1$ and $\G_2$ play the role of the cores
of macroscopic ``zero energy excitations''~\cite{KM00} that flip as we go from
one pure state to another. $\G_1$ separates state space into  
$\C$ and $\bar\C$. Within $\C$, $\G_2$ induces  a further non-trivial
separation of the states, into clusters $\C_1$ and $\C_2$. 
Each cluster $\C_\alpha$ contains one or more pure states.  
When $\G_2$ is large  (``macroscopic''), a pair of states $\mu \in \C_1$
and $\nu \in \C_2$ will belong to different pure states, and their
overlap $q^{\mu \nu}$ will contribute to $P_J^o(q)$.
Hence, we define a new distribution ${\tilde P}_J^o(q)$, to which
only pairs of states $\mu \in \C_1$ and $\nu \in \C_2$  contribute.
This function is a lower bound to $P^o_J(q)$, since we might have for some
realizations a third macroscopic cluster, in which case $\C_1$ contains states
from more than one pure state. When this happens, some pairs of states, 
both taken from $\C_1$, contribute to $P^o_J(q)$, and we do not include them in 
${\tilde P}_J^o(q)$.
In order to assure that  $\C_1$ and $\C_2$ indeed  do not belong to
one pure state, we consider only those realizations for which $|\G_2|>0.05N$.
Otherwise, we set $\tilde P_J^o(q)=0$.

The method we used to partition the states was based on a clustering procedure.
It is important to stress the fact that the main result of the present study, 
that there are
states whose overlap contribution should be separated from the self-overlap
peak and does not vanish in the thermodynamic limit, does {\it not} depend
qualitatively on the way the state clusters are determined. 
In fact, 
any method, which projectes out a particular contribution to $P(q)$ and
has a nonvanishing weight in the $L \rightarrow \infty$ limit, 
will lead to the same conclusion. The only
requirements are  that the method is applied for all system sizes in the
same way and the contribution is measured in absolute weights with respect to
the total $P(q)$.

We determined~\cite{PRL} for each $L$ the 
size distributions $|\G_2|/N$ and found that they are nearly the same for 
$4 \leq	L \leq 8$, indicating convergence. $|\G_2|$ scales as $N=L^3$ ; 
for $L=6$ the average value of $|\G_2|/N$  is 0.07 and its standard
deviation 0.09; for $L=8$ the numbers are 0.08 and 0.10, respectively.

Since the limiting size distribution of $\G_2$ is non-trivial, 
we expect a non-trivial
$\tilde P^o_J(q)$ 
as long as $\C_2$ does not vanish.
The size $|\C_2|$ of this state cluster 
is determined by the correlation between
the spins of $\G_1$ and $\G_2$. If this correlation approaches 1, this means
that $\G_2$ has a low probability to flip without $\G_1$, resulting in
$|\C_1|\gg|\C_2|$. The average correlation between these domains is
\begin{equation}
\bar c_{12} = {1\over|\G_1||\G_2|} 
\sum_{i\in\G_1} \sum_{j\in\G_2} {c_{ij}}^2 \;,
\end{equation}
where $c_{ij}=\langle S_i S_j \rangle$ is the correlation between spins $i$
and $j$. The weight of the contribution to $P(q)$ by pairs of states in
which $\G_2$ is flipped without $\G_1$ or vice versa can be evaluated
\cite{PRL} by $(1- \bar c_{12})/2$.
We found~\cite{PRL} that $\bar c_{12}$ does not extrapolate to 1
as $L \rightarrow \infty$. For a realization $\{J\}$, in which  
$\bar c_{12}<1$
and $|\G_2|>0$, the function $P_J^o(q)$ will be non-trivial, i.e. it will have
a finite support for $-1<q<1$. 

\begin{figure}[t]
\centerline{\psfig{figure=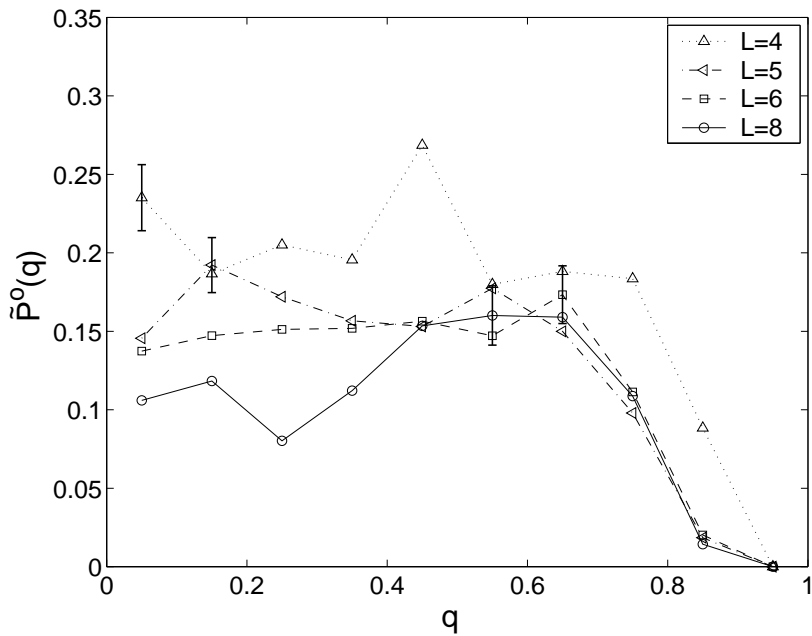,width=11cm}}\vspace{2mm}
\caption{The partial distribution $\tilde P^o(q)$ for $L=4,5,6,8$.
It is normalized so that $2\int_0^1 \tilde P^o(q) dq$ is its weight in
the total $P(q)$. For each $L$ the largest error bar is shown.}
\label{fig:pqo}
\end{figure}

\begin{figure}[t]
\centerline{\psfig{figure=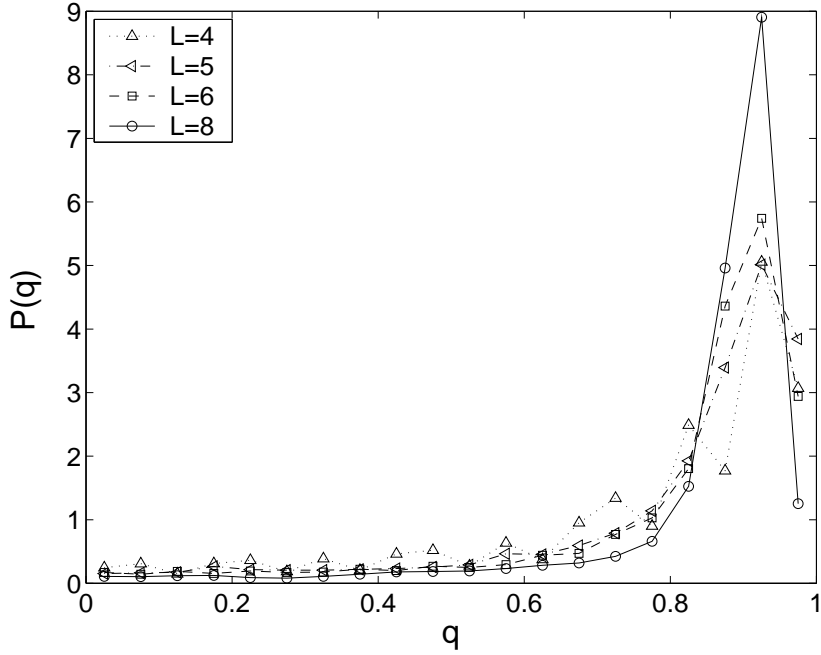,width=11cm}}\vspace{2mm}
\caption{The distribution $P(q)$ for $L=4,5,6,8$. The size of the
error bars is of the order or less than the size of the symbols.}
\label{fig:pq}
\end{figure}

To show explicitly that this indeed is the case, we studied $\tilde P^o(q)$ and
$P(q)$.
The function $\tilde P^o(q) = [\tilde P_J^o(q)]_J$, presented in Fig.
\ref{fig:pqo}, is a conservative estimate (and a lower bound) for
$P^o(q)$. It also has a clear physical meaning. $\tilde P^o(q)$ is the
distribution of overlaps between pairs of states on the two sides of the
second largest free energy barrier in the system.
For comparison, we also present the full $P(q)$ in Fig \ref{fig:pq}.

${\tilde P}^o(q)$ has low values and large relative errors
for $q <0.4$. In this range its values decrease with increasing $L$.
On the other hand,  in the interval
$0.4 \leq q \leq 0.7$ it seems to have converged.
Therefore we chose this range for our analysis, and calculated
the integrals $x^*$ (see eq. (\ref{eq:x*})) and
\begin{equation}
\tilde x^{*o} = 2\int_{0.4}^{0.7} {\tilde P}^o(q) dq \;.
\label{eq:xto*}
\end{equation}
The values obtained for $L=4,5,6,8$ are presented in Table \ref{tab:x}.
Perhaps the most direct evidence for our claim is the manner in which
the values of $\tilde x^{*o}$ level off as the size increases, at 0.047. 
On the other hand,
those of $x^*$ decrease with size. We performed a fit of the latter to the form
\begin{equation}
x^* = A + B L^{-y}
\end{equation}
The results of several attempts to fit the data to this form are summarized in
Table \ref{tab:fit}. 
The best fit (with $\chi^2=1.0 \times 10^{-5}$) 
was obtained for $y=2.06(49)$ and $A=0.042(15)$, 
which is close to $0.047$.
Imposing this value, i.e. setting $A=0.047$ and fitting $B$ and $y$,
we had a somewhat larger $\chi^2=1.1 \times 10^{-5}$; imposing $A=0$ yields a 
worse fit, with 
$\chi^2=4.2 \times 10^{-5}$.
We believe that these results clearly show that $P(q)$ is non-trivial.
\begin{table}[ht]
\begin{center}
\begin{tabular}[t]{|c|c|c|c|c|}
$L$ & $x^*$ & $\tilde x^{*o}$ & $x_{1/2}$  & $\tilde x_{1/2}^o$ \\ \hline
4 & 0.161(5) & 0.064(10) & 0.157(7) & 0.109(14) \\
5 & 0.115(5) & 0.048(11) & 0.105(6) & 0.082(15) \\
6 & 0.096(5) & 0.048(10) & 0.095(5) & 0.074(14) \\
8 & 0.070(4) & 0.047(12) & 0.062(4) & 0.057(14) \\ 
\end{tabular}
\end{center}
\caption{Values of the observables (defined in eq. (\ref{eq:xa}),
(\ref{eq:xtilde}), (\ref{eq:x*}) and (\ref{eq:xto*})) for different
system sizes.}
\label{tab:x}
\end{table}

To make contact with previous analysis we also calculated $x_{1/2}$ and
performed similar fits, the results of which are also presented in Table II. 
As discussed above, in this range of $q$ the
function ${\tilde P}^o(q)$ has larger statistical fluctuations, and 
is decreasing with size (to a limiting value that is  expected
to be small, albeit non-zero). Indeed the best 
fit for $x_{1/2}$ is attained for $y=2.07(1.51)$ and $A=0.036(47)$, 
with $\chi^2=1.0 \times 10^{-4}$.
Note that this $\chi^2$ is 10 times the value obtained when fitting $x^*$.
Since our estimated  value of $A$, as well as the estimates of others 
\cite{Berg94,Hartmann3d} is much smaller then the values of $x_{1/2}$ used
to perform the fit,  it is hard to distinguish, by means of  
this extrapolation,
between $A=0$ and a small positive $A$. Indeed, when we impose $A=0$ or both
$A=0, B=1$ we get fits of comparable quality, with an exponent which is
consistent with Hartmann's estimate.

Finally, we attempted to fit the data for $\tilde x^o_{1/2}$.
The values of $\tilde x^o_{1/2}$
for the system sizes used are smaller and noisier the the results for $x^*$. 
Nevertheless, using the same fit for
$\tilde x^o_{1/2}$ yields minimum of $\chi^2=1.5 \times 10^{-5}$ for 
$y=1.94(1.02)$
and $A=0.040(20)$, quite consistent with the results obtained for $x^*$.

The authors thank A.P. Young and M. Palassini for most helpful correspondence.
AKH was supported by the Graduiertenkolleg
``Modellierung und Wissenschaftliches Rechnen in
Mathematik und Naturwissenschaften'' at the
{\em In\-ter\-diszi\-pli\-n\"a\-res Zentrum f\"ur Wissenschaftliches
  Rechnen} in Heidelberg and the
{\em Paderborn Center for Parallel Computing}
 by the allocation of computer time.  AKH obtained financial
 support from the DFG ({\em Deutsche Forschungs Gemeinschaft}) under
 grant Zi209/6-1. ED and GH were supported by the Germany-Israel Science
 Foundation (GIF).

\begin{table}[h]
\begin{center}
\begin{tabular}[t]{|l|llll|}
fit                     &  $A$  &  $B$ &  $y$ &$\chi^2$\\ \hline
$x^*$, best        & 0.042(15) &2.06(1.16)& 2.06(49) &
$1.0\times10^{-5}$\\
$x^*$, imposed $A$ & 0.047     & 2.45(36) & 2.21(10) &
$1.1\times10^{-5}$\\
$x^*$, imposed $A$ & 0         & 0.88(13) & 1.24(9)  &
$4.2\times10^{-5}$\\
\hline
$x_{1/2}$, best    & 0.036(47) &2.10(3.68)&2.07(1.51)&
$1.0\times10^{-4}$\\
$x_{1/2}$, imposed $A$  & 0    & 0.97(26) & 1.33(17) &
$1.3\times10^{-4}$\\
$x_{1/2}$, imposed $A,B$& 0    & 1        & 1.35(2)  &
$1.3\times10^{-4}$\\
\hline
$\tilde x_{1/2}^o$,
best&0.040(20)&1.00(1.16)&1.94(1.02)&$1.5\times10^{-5}$\\
\end{tabular}
\end{center}
\caption{Best fit parametrs for $x(L)=A+B L^{-y}$.}
\label{tab:fit}
\end{table}




\end{document}